\documentclass[runningheads,a4paper]{llncs}
\usepackage{amssymb,amsmath,amscd,amstext,graphicx,texdraw,mathdots,gastex,comment}
\usepackage{color}
\newcommand{\N}{\mathbb{N}}

\newcommand{\gk}{\kappa}

\newcommand{\gl}{\lambda}

\newcommand{\mc}{\mathcal}
\newcommand{\bx}{\mathbin{\Box}}

\newcommand{\qq}{\qquad}
\newcommand{\lma}{\left(\begin{matrix}}
\newcommand{\rma}{\end{matrix}\right)}

\newcommand{\bm}{\begin{matrix}}
\newcommand{\enm}{\end{matrix}}
\newcommand{\bc}{\begin{center}}
\newcommand{\ec}{\end{center}}
\newenvironment{keywords}{
       \list{}{\advance\topsep by0.35cm\relax\small
       \leftmargin=1cm
       \labelwidth=0.35cm
       \listparindent=0.35cm
       \itemindent\listparindent
       \rightmargin\leftmargin}\item[\hskip\labelsep
                                     \bfseries Keywords:]}
     {\endlist}
\begin{document}

\title{$k$-Colorability is Graph Automaton Recognizable}
\author{Antonios Kalampakas}
\institute{Department of Production Engineering and Management,
Laboratory of Computational Mathematics,
School of Engineering, Democritus University of Thrace,
V.Sofias 12, Prokat, Building A1, 67100Xanthi, Greece, akalampakas@gmail.com}
\maketitle
\begin{abstract}
Automata operating on general graphs have been introduced by virtue of graphoids. In this paper we construct a graph automaton that recognizes $k$-colorable graphs.
\end{abstract}


\section{Introduction}

Automata on general (hyper)graphs were constructed for the first time in \cite{BK3} by utilizing the algebraic properties of graphoids, i.e., magmoids  satisfying the 15 equations of graphs which are specified in \cite{BK1}. The notion of magmoids, introduced by Arnold and Dauchet in \cite{AD}, is the algebraic structure employed to generate graphs from a finite set in a role similar to that of monoids for the generation of strings. A magmoid is a doubly ranked set equipped with two operations which are associative, unitary and mutually coherent in a canonical way. In \cite{EV} Engelfriet and Vereijken proved that every graph with edges labeled  over a finite doubly ranked set $\Sigma$ can be built from a specific finite set of elementary graphs $D$, together with the elements of $\Sigma$, by using the operations of graph product and graph sum. From this result it is derived that graphs can be organized into a magmoid with operations product and sum. Although, as it was observed in \cite{EV}, every hypergraph can be constructed in an infinite number of ways, this ambiguity was settled in \cite{BK1} by determining a \emph{finite} set of equations $\mathcal{E}$ with the property that two expressions represent the same hypergraph, if and only if, one can be transformed into the other
through them.

Commencing from this result, a \emph{graphoid} $\mathbf{M}$ is defined as a magmoid with a designated set of elements that satisfy the equations $\mathcal{E}$. Hence $GR(\Sigma)$ can be structured into a graphoid by virtue of the set $D$ of elementary graphs. The \emph{relational magmoid} over a set $Q$ is constructed by defining the operations of composition and sum on the set of relations from $Q^m$ to $Q^n$. This set is structured into a \emph{relational graphoid} over $Q$, by specifying a set $D$ of relations that satisfy the equations $\mathcal{E}$. A relational graphoid is called abelian when a particular relation of $D$ consists of all the transpositions in $Q$.
In \cite{BK3} graph automata, with state set $Q$, were introduced by virtue of a specific abelian relational graphoid, denoted here by $\mathbf{TSRel(Q)}$, and by exploiting the fact that $GR(\Sigma)$ is the free graphoid generated by $\Sigma$. In the same paper it is postulated that different kinds of graphoids will produce graph automata with diverse operation and recognition capacity. In \cite{Kal2} it is shown that all abelian relational graphoids are characterized in the following way: a set $Q$ generates an abelian relational graphoid if and only if $Q$ is partitioned  into  disjoint abelian groups with operations univocally correlated with $D$. In other words it is proved that organizing $Rel(Q)$ in a relational graphoid is equivalent to partitioning the set $Q$ and structuring every class in a group.

The particular graphoid $\mathbf{TSRel(Q)}$ employed in \cite{BK3} corresponds to the partitioning of $Q$ into singleton sets each one being the trivial group. Hence, due to \cite{Kal2}, it is the simplest possible abelian relational graphoid. In this paper we construct a graph automaton over $\mathbf{TSRel(Q)}$ which accepts the $k$-colorable graphs, $k\geq 1$. Hence it is manifested that graph automata are capable of recognizing important classes of graphs even when operating on the most trivial of the known graphoids. In the following section we recall basic definition for magmoids and hypergraphs. Graphoids are presented in Section \ref{S:3} and the definition of a graph automaton is given. In Section \ref{S:4} we construct the graph automaton that recognizes $k$-colorable graphs and illustrate its operation. Conclusions and future work are discussed in the last section.

\section{Magmoids and Hypergraphs}

A doubly ranked set  $(A_{m,n})_{m,n\in\N}$ is a set $A$ together with a function $rank: A\to \N\times \N$ we set $A_{m,n}=\{a \in A \mid
rank(a)=(m,n)\}$. In what follows we will drop the subscript   and denote a doubly ranked set simply by $A=(A_{m,n})$.
A \emph{magmoid} is a doubly ranked set $M=(M_{m,n})$
equipped with two operations
\begin{center}
$\circ : M_{m,n}\times M_{n,k}\to M_{m,k}$,
\quad
$\bx : M_{m,n}\times M_{m',n'}\to M_{m+m',n+n'}$,
\end{center}
which are associative in the obvious way, satisfy the
distributivity law
\[
(f\circ g)\bx (f'\circ g')=(f\bx f')\circ (g\bx g')
\]
whenever all the above operations are defined and are equipped with a
sequence of constants $e_n\in M_{n,n}$,  called units, such that
\[
e_m\circ f=f=f\circ e_n,\;\; e_0\bx f=f=f\bx e_0,\; e_m\bx e_n =e_{m+n}
\]
hold for all $f\in M_{m,n}$ and  all $m,n\in \N$.
Notice that, due to the last equation, the elements $e_n$ are uniquely determined by $e_1$. From now on $e_1$ will be
simply denoted by $e$. The free magmoid $mag(\Sigma)$ generated by a doubly ranked set $\Sigma$ is constructed in \cite{BK1}. The sets $Rel_{m,n}(Q)$ of all relations from $Q^m$ to $Q^n$
\[
Rel_{m,n}(Q)=\{R\mid R \subseteq Q^m\times Q^n \}
\]
can be structured into a magmoid with $\circ$ being the usual
relation composition and $\bx$ defined as follows: for $R\in Rel_{m,n}(Q)$ and $S\in
Rel_{m',n'}(Q)$
\[
R\bx S=\{(u_1u_2,v_1v_2)\mid (u_1,v_1)\in R\mbox{ and }(u_2,v_2)\in S)\},
\]
where $u_1\in Q^m$, $u_2\in Q^{m'}$, $v_1\in Q^n$, $v_2\in Q^{n'}$. Notice that $Q^0=\{\varepsilon\}$, where $\varepsilon$ is the empty word of $Q^*$. The units are given by
$ e_0=\{(\varepsilon,\varepsilon)\}$ and $
e=\{(g,g)\mid g\in Q\}$.
We denote by $Rel(Q)=(Rel_{m,n}(Q))$ the magmoid  constructed in this way and call it the \emph{relational magmoid of} $Q$.

 An $(m,n)$-(hyper)graph $G=(V,E,s,t, l,begin,end)$ with edge labels from a doubly ranked set $\Sigma =(\Sigma_{m,n})$ is a tuple   consisting of the set of nodes or vertices $V$, the set of edges $E$,   the source and target functions $s:E \to V ^+$ and $t:E\to V ^+$ respectively,    the labeling function $l:E \to \Sigma$ such that  $rank(l(e))=(|s(e)|,|t(e)|)$, for all $e\in E $,  and  the sequences of begin and end nodes $begin\in V^*$ and $end\in V^*$ with $|begin|=m$ and $|end|=n$. Notice that vertices can be duplicated in the begin and end sequences of the graph and also at the sources and targets of the edges. Isomorphism between two graphs is defined in the obvious way and we shall not distinguish between two isomorphic  graphs. The set of all  $(m,n)$-graphs over $\Sigma$ is
denoted by $GR_{m,n}(\Sigma)$  and we set $GR(\Sigma)=(GR_{m,n}(\Sigma))_{m,n\in\N}$. Ordinary unlabeled directed graphs are obtained as a special case of hypergraphs i.e., in the case that each hyperedge is binary (has one source and one target), every edge has the same label and the sequences $begin$ and $end$ are the empty word.

If $G$ is the $(m,n)$-graph  $(V,E,s,t,l,begin,end)$ and $H$ is the $(n,k)$-graph  $(V',E',s',t',l',begin',end')$ then their \emph{product} $G\circ
H$ is the $(m,k)$-graph that is obtained by taking the disjoint union of $G$ and $H$ and then identifying
the $i^{th}$ end node of $G$ with the $i^{th}$ begin node of $H$, for every $i\in \{1,...,n\}$; also, $begin(G\circ H)=begin(G)$ and
$end(G\circ H)=end(H)$. The \emph{sum} $G \bx H$ of arbitrary graphs $G$ and $H$ is their disjoint union with their sequences of begin nodes concatenated and similarly for their end nodes (see \cite{BK3,Kal2} for examples). For every $n\in \mathbb{N}$ we denote by $E_n$ the discrete graph
of rank $(n,n)$ with nodes $x_1,...,x_n$ and $begin =end =x_1\cdots x_n$; we write $E$ for $E_1$.  It is straightforward to verify that $GR(\Sigma )=(GR_{m,n}(\Sigma))$ with the operations  defined above is a magmoid, whose units are the graphs $E_n$.

\section{Graphoids and Graph Automata}\label{S:3}

Now we present graph automata by employing the algebraic structure of graphoids as introduced in \cite{BK3}. We denote by $I_{p,q}$ the discrete $(p,q)$-graph that has a single node $x$ and whose begin and end sequences are $x\cdots x$ ($p$ times) and $x\cdots x$ ($q$ times) respectively, $\Pi$ is the discrete $(2,2)$-graph that has two nodes $x$ and $y$ and whose begin and end sequences are $xy$ and $yx$, respectively, also for every $\sigma\in\Sigma_{m,n}$, we denote again by $\sigma$ the $(m,n)$-graph having only one edge and $m+n$ nodes $x_1,\dots ,x_m,y_1,\dots ,y_n$. The edge is labeled by $\sigma$, and the begin (resp. end sequence) of the graph is the sequence of sources (resp. targets) of the edge, viz. $x_1\cdots x_m$ (resp.
$y_1\cdots y_n$).

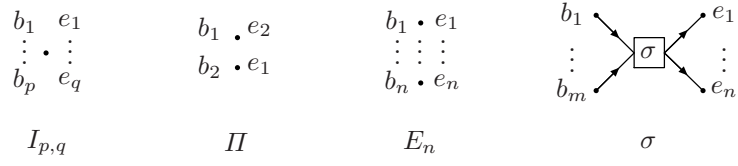
\vspace{4mm}

\begin{minipage}[t]{2cm}

\begin{center}
\begin{texdraw}
\drawdim cm \linewd 0.021 \arrowheadtype t:F \arrowheadsize l:0.12
w:0.10 \move(2.4 2)\fcir f:0 r:0.03

\textref h:C v:C  \htext(2.4 0.8){$I_{p,q}$} \htext(2.1
2.4){$b_1$} \htext(2.1 2.1){$\vdots$} \htext(2.1 1.6){$b_p$}
\htext(2.7 2.4){$e_1$} \htext(2.7 2.1){$\vdots$} \htext(2.7
1.6){$e_q$}

\end{texdraw}
\end{center}
\end{minipage}
\quad
\begin{minipage}[t]{2cm}
\begin{center}
\begin{texdraw}

\drawdim cm \linewd 0.021 \arrowheadtype t:F \arrowheadsize l:0.12
w:0.10 \move(4 2.2)\fcir f:0 r:0.03 \move(4 1.8)\fcir f:0 r:0.03

\textref h:C v:C  \htext(4 0.8){$\Pi$} \htext(3.65 2.3){$b_1$}
\htext(3.65 1.8){$b_2$} \htext(4.3  2.3){$e_2$} \htext(4.3
1.8){$e_1$}

\end{texdraw}
\end{center}
\end{minipage}
\quad
\begin{minipage}[t]{2cm}
\begin{center}
\begin{texdraw}

\drawdim cm \linewd 0.021 \arrowheadtype t:F \arrowheadsize l:0.12
w:0.10 \move(5.6 2.4)\fcir f:0 r:0.03 \move(5.6 1.6)\fcir f:0
r:0.03

\textref h:C v:C  \htext(5.6 0.8){$E_n$} \htext(5.3 2.4){$b_1$}
\htext(5.3 2.1){$\vdots$} \htext(5.3 1.6){$b_n$} \htext(5.95
2.4){$e_1$} \htext(5.9 2.1){$\vdots$} \htext(5.6 2.1){$\vdots$}
\htext(5.95 1.6){$e_n$}

\end{texdraw}
\end{center}
\end{minipage}
\quad
\begin{minipage}[t]{3cm}
\begin{center}
\begin{texdraw}

\drawdim cm \linewd 0.021 \arrowheadtype t:F \arrowheadsize l:0.12
w:0.10      \move (7.3 1.5)\fcir f:0 r:0.03 \lvec(7.8 2)\move (7.3
1.5)\avec(7.6 1.8) \move (7.3 2.5)\fcir f:0 r:0.03 \lvec(7.8
2)\move(7.3 2.5) \avec(7.6 2.19)\move(7.8 2)

\move(7.8 1.8)\lvec(7.8 2.2)\lvec(8.2 2.2)\lvec(8.2 1.8)\lvec(7.8
1.8)\move(8.2 2)\lvec(8.7 2.5)\fcir f:0 r:0.03 \move(8.2
2)\lvec(8.7 1.5)\fcir f:0 r:0.03 \move(8.2 2)\avec(8.5
2.3)\move(8.2 2)\avec(8.5 1.7)

\textref h:C v:C  \htext(8 2){$\sigma$}\htext(9
2.5){$e_1$}\htext(9 2){$\vdots$}\htext(9 1.5){$e_n$}\htext(7
2.5){$b_1$}\htext(7 1.5){$b_m$} \htext(7 2){$\vdots$}\htext(8
0.8){$\sigma$}
\end{texdraw}
\ec

\end{minipage}

\vspace{4mm}

Engelfriet and Vereijken, in \cite{EV},  presented an algorithm that inductively constructs every graph $G\in GR(\Sigma)$ from the set $\Sigma\cup\{\Pi,I_{01},I_{21},I_{10},I_{12},\}$ by using graph product and graph sum.  However, there are infinitely many ways to construct a given graph. This was overridden  by identifying a finite set $\mathcal{E}$ of equations  with the property that two expressions represent the same graph if and only if one can be transformed into the other through these equations \cite{BK1}. It is evident from this discussion that the equations $\mathcal{E}$ are satisfied in $GR(\Sigma)$. Magmoids with such a property are called \emph{graphoids}. Formally, a graphoid $\mathbf{M}=(M,D)$ consists of a magmoid $M$ and a set $D=\{s,d_{01},d_{21},d_{10},d_{12}\}$, with $s\in M_{2,2}$, $d_{\gk\gl}\in M_{\gk,\gl}$,  such that the following equations hold:\\
\begin{minipage}[b]{70pt}
\begin{align}\label{E:3} s\circ s =e_2   \end{align}  \end{minipage}  \begin{minipage}[b]{270pt}\begin{align}\label{E:4}  \left(s \bx e \right) \circ \left( e \bx
s \right)\circ \left( s \bx e \right)   = \left( e \bx s
\right) \circ \left( s \bx e \right) \circ \left( e \bx s
\right)
\end{align}
\end{minipage}\\
\begin{minipage}[b]{170pt}
\begin{align}\label{E:5}
\left( e \bx d_{21}\right)\circ d_{21}=\left(
d_{21}\bx e \right)\circ d_{21} \end{align}  \end{minipage}\qq\quad  \begin{minipage}[b]{140pt}\begin{align}\label{E:6}    \left(e \bx d_{01}
\right)\circ d_{21}=e
\end{align}
\end{minipage}\\
\begin{minipage}[b]{120pt}
\begin{align}\label{E:7}
s\circ d_{21}=d_{21}\end{align}  \end{minipage}\qq\quad  \begin{minipage}[b]{180pt}\begin{align}\label{E:8}  \left( e\bx d_{01}
\right)\circ s = \left(
 d_{01}\bx e\right)
\end{align}
\end{minipage}\\
\begin{minipage}[r]{300pt}
\begin{align}\label{E:9}
\left(
s \bx e )\circ ( e \bx s \right)\circ\left( d_{21}\bx e
\right)= \left( e\bx d_{21}\right)\circ s
\end{align}
\end{minipage}\\
\begin{minipage}[b]{170pt}
\begin{align}\label{E:10}   d_{12}\circ \left(  e\bx d_{12} \right)=d_{12}\circ
\left( d_{12}\bx e \right)\end{align}  \end{minipage} \qq  \begin{minipage}[b]{140pt}\begin{align}\label{E:11}   d_{12}\circ \left( e\bx
d_{10} \right)=e
\end{align}
\end{minipage}\\
\begin{minipage}[b]{120pt}
\begin{align}\label{E:12}
 d_{12}\circ s =d_{12} \end{align}  \end{minipage} \qq  \begin{minipage}[b]{180pt}\begin{align}\label{E:13}   s\circ \left(
e\bx d_{10} \right)= \left( d_{10}\bx e \right)
\end{align}
\end{minipage}\\
\begin{minipage}[b]{300pt}
\begin{align}\label{E:14}
 \left( d_{12}\bx e \right)\circ (  e \bx s )\circ (s \bx
e) = s\circ \left( e\bx d_{12} \right)
\end{align}
\end{minipage}\\
\begin{minipage}[b]{100pt}
\begin{align}\label{E:15}
 d_{12}\circ d_{21}=e \end{align}  \end{minipage} \qq\quad  \begin{minipage}[b]{200pt}\begin{align}\label{E:16} \left( d_{12}\bx e
\right)\circ \left(e\bx d_{21} \right)= d_{21}\circ d_{12}
\end{align}
\end{minipage}\\
\begin{minipage}[b]{340pt}
\begin{align}\label{E:17}
 s_{m,1}\circ (p \bx e ) = (e \bx p )\circ
s_{n,1},\quad\mbox{ for all } p\in M_{m,n}
\end{align}
\end{minipage}\\ \\
where $s_{m,1}$ is defined inductively by $s$  and represents the graph associated with the permutation
that interchanges the last $n$ numbers with the first one \cite{BK1}. We point out that although (\ref{E:17}) is a set of equations it only has to be valid for the elements of $\Sigma$ in order to hold for every element of a magmoid generated by $\Sigma$ \cite{BK1}.   Thus the pair $\mathbf{GR(\Sigma)}=(GR(\Sigma),D)$, with $D=\{\Pi ,I_{01},I_{21}, I_{10}, I_{12}\}$ is a graphoid and in fact it is the free graphoid
generated by $\Sigma$ as it is illustrated in \cite{BK3}. Given graphoids $(M,D)$ and $(M',D')$, a magmoid morphism $H:M\to M'$ preserving $D$-sets, i.e.,  $H(s)=s'$ and $H(d_{\gk\gl})= d'_{\gk\gl}$, is called a morphism of graphoids.

Graphoids constructed from the magmoid of relations $Rel(Q)$ over a given set $Q$ are called \emph{relational graphoids} and a relational graphoid is called abelian when  $s=\{(g_1g_2,g_2g_1)\mid g_1,g_2\in Q\}$. The abelian relational graphoid $\mathbf{TSRel(Q)}=(Rel(Q),D)$ that was used for the introduction of graph automata is constructed by setting $s$ as above and
  $d_{01}=\{(\varepsilon,g)\mid
g\in Q\}$,
   $d_{21}=\{(gg,g)\mid
g\in Q\}$,
$d_{10}=\{(g,\varepsilon)\mid
g\in Q\}$,
 $d_{12}=\{(g,gg)\mid
g\in Q\}$.

A \emph{nondeterministic relational graph automaton} is a structure
\bc
$\mathcal{A}=(\Sigma,Q,\bold{Rel}(Q),\delta ,I ,T )$,
\ec
where $\Sigma$ is the doubly ranked set of hyperedge labels,     $Q$ is the finite set of states,   $\bold{Rel}(Q)$ is a relational graphoid over $Q$,    $\delta  :\Sigma \to \bold{Rel}(Q)$  is the doubly ranked transition function and  $I ,T $ are initial and final rational subsets of
$Q^*$. The function $\delta$ is uniquely extended into a morphism of graphoids
$\bar{\delta} :GR(\Sigma)\to \bold{Rel}(Q)$, where $\bar{\delta} (I_{\gk\gl})=d_{\gk\gl}$ and $\bar{\delta} (\Pi)=s$,
and the behavior of $\mathcal{A}$ is given by
\[
|\mathcal{A}|=\{F\mid F\in GR_{m,n}(\Sigma),\;
\bar{\delta}_\mathcal{A}(F)\cap (I_{\mc{A}}^{(m)}\times
T_{\mc{A}}^{(n)})\neq \emptyset,\; m,n\in \N\}
\]
where $I_{\mc{A}}^{(m)}=I_\mathcal{A}\cap Q^m$ and
$T_{\mc{A}}^{(n)}=T_\mathcal{A}\cap Q^n$. From their construction, graph automata are
finite machines due to the fact that the set of equations
$(\ref{E:3})$-$(\ref{E:17})$ is finite.
A graph language is called recognizable whenever it is obtained as
the behavior of a graph automaton. The class of all such languages
over the doubly ranked set $\Sigma$ is denoted by
 $Rec(\Sigma)$.

\section{A Graph Automaton Recognizing $k$-colorable Graphs}\label{S:4}

In this section we construct a relational graph automaton over the abelian graphoid $\mathbf{TSRel(Q)}$ recognizing $k$-colorable graphs.

For $k\in \N^*$, we set $\mc{A}^k_{clr}=(\Sigma,Q,\mathbf{TSRel(Q)},\delta,I,T)$ with
\begin{itemize}
  \item $\Sigma =\{a\}$, $rank(a)=(1,1)$,
  \item $Q=\{1,2,\dots ,k\}$,
  \item $\delta (a)=\{(i,j)\mid i,j\in Q, i\neq j\}$,
  \item $I=T=\{\varepsilon\}$.
\end{itemize}
It is clear that the automaton $\mc{A}^k_{clr}$ reads unlabeled $(0,0)$-graphs with binary edges (one source and one target per edge), i.e., ordinary directed graphs.

As an example we shall illustrate the operations of $\mc{A}^2_{clr},\mc{A}^3_{clr}$ and $\mc{A}^4_{clr}$ on the following graphs, where the label in every edge is $a$ and thus omitted.
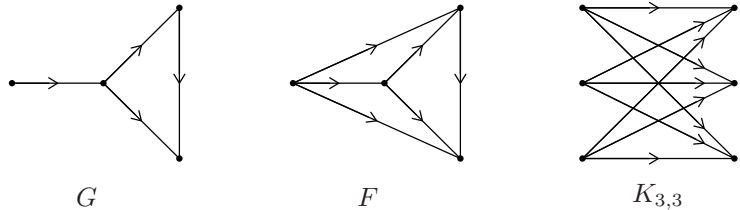
\bc
\begin{minipage}[b]{300pt}
\bc
\begin{texdraw}
\drawdim mm \linewd 0.21 \arrowheadtype t:H \arrowheadsize l:1.2
w:1.4

\move(42 10)\fcir f:0 r:0.4 \lvec
(32 20) \move (32 20) \avec (37 15)  \move(42 30)\fcir f:0 r:0.4 \lvec
(32 20)  \move (32 20) \avec (37 25) \move(42 30)  \avec (42 20)\move (42 30)\lvec
(42 10) \move(20 20)\fcir f:0 r:0.4  \avec (26 20)\move (20 20)\lvec
(32 20) \fcir f:0 r:0.4

\textref h:C v:C   \htext(30 5){$G$}
\end{texdraw}\qq\qq
\begin{texdraw}
\drawdim mm \linewd 0.21 \arrowheadtype t:H \arrowheadsize l:1.2
w:1.4

\move(42 10)\fcir f:0 r:0.4 \lvec
(32 20) \move (32 20) \avec (37 15)  \move(42 30)\fcir f:0 r:0.4 \lvec
(32 20)  \move (32 20) \avec (37 25) \move(42 30)  \avec (42 20)\move (42 30)\lvec
(42 10) \move(20 20)\fcir f:0 r:0.4  \avec (26 20)\move (20 20)\lvec
(32 20) \fcir f:0 r:0.4  \move (20 20) \lvec (42 30) \move (20 20) \avec (31 25)
\move (20 20) \lvec (42 10) \move (20 20) \avec (31 15)

\textref h:C v:C   \htext(30 5){$F$}
\end{texdraw} \qq\qq
\begin{texdraw}
\drawdim mm \linewd 0.21 \arrowheadtype t:H \arrowheadsize l:1.2
w:1.4

\move(10 10)\fcir f:0 r:0.4
\lvec (30 10) \move (10 10) \avec (20 10)  \move (10 10)
\lvec (30 20) \move (10 10)  \avec (26 18)  \move (10 10)
\lvec (30 30)  \move (10 10) \avec (26 26)

\move(10 20)\fcir f:0 r:0.4
\lvec (30 10) \move (10 20)  \avec (26 12)  \move (10 20)
\lvec (30 20) \move (10 20)   \avec (26 20) \move (10 20)
\lvec (30 30)  \move (10 20)  \avec (26 28)

\move(10 30)\fcir f:0 r:0.4
\lvec (30 10) \move (10 30)  \avec (26 14)  \move (10 30)
\lvec (30 20) \move (10 30)  \avec (26 22)  \move (10 30)
\lvec (30 30)  \move (10 30) \avec (20 30)

\move(30 10)\fcir f:0 r:0.4
\move(30 20)\fcir f:0 r:0.4
\move(30 30)\fcir f:0 r:0.4

\textref h:C v:C   \htext(20 5){$K_{3,3}$}
\end{texdraw}
\ec
\end{minipage}
\ec
One of the representations of $G$ is
\[
G=I_{01}\, a\, I_{12}\lma a\\ a\rma \lma a\\ E \rma I_{21} \, I_{10}
\]
where graph product and graph sum are denoted, for simplicity, by horizontal and vertical concatenation. Note that since $GR(\Sigma)$ is the free graphoid, the operation of the automaton is the same for every representation of $G$. The consumption of $G$ by $\mc{A}^3_{clr}$ gives
\[
\bar{\delta}(G)=d_{01}\, \delta(a)\, d_{12}\lma \delta(a)\\ \delta(a)\rma \lma \delta(a)\\ e \rma d_{21} \, d_{10}
\]
and an accepting sequence of transitions is
\[
\{\varepsilon\}d_{01}\{1\}\delta(a)\{2\} d_{12} \left\{\begin{matrix} 2 \\ 2\end{matrix}\right\} \lma \delta(a)\\ \delta(a)\rma \left\{\begin{matrix} 3 \\ 1\end{matrix}\right\}  \lma  \delta(a)\\ e \rma \left\{\begin{matrix} 1 \\ 1\end{matrix}\right\} d_{21}\{1\} \, d_{10}\{\varepsilon\}
\]
where the states are indicated in brackets. In graphical representation the states that the automaton reaches at each state are
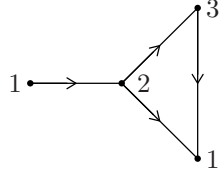
\bc
\begin{minipage}[b]{300pt}
\bc
\begin{texdraw}
\drawdim mm \linewd 0.21 \arrowheadtype t:H \arrowheadsize l:1.2
w:1.4

\move(42 10)\fcir f:0 r:0.4 \lvec
(32 20) \move (32 20) \avec (37 15)  \move(42 30)\fcir f:0 r:0.4 \lvec
(32 20)  \move (32 20) \avec (37 25) \move(42 30)  \avec (42 20)\move (42 30)\lvec
(42 10) \move(20 20)\fcir f:0 r:0.4  \avec (26 20)\move (20 20)\lvec
(32 20) \fcir f:0 r:0.4

\textref h:C v:C   \htext(18 20){$1$} \htext(35 20){$2$} \htext(44 10){$1$} \htext(44 30){$3$}
\end{texdraw}
\ec
\end{minipage}
\ec
which is actually a proper 3-coloring of $G$. Similarly, a representation for $F$ is
\[
F=I_{01}\,I_{13} \lma a\\ a\\a\rma\,\lma E\\ I_{12}\\ E \rma \lma E\\a\\ a\\E\rma \lma I_{21}\\ I_{21}\rma \lma a\\ E \rma I_{21} \, I_{10}
\]
and it is clear that there exists no successful transition of $F$ by $\mc{A}^3_{clr}$. A successful transition of $F$ by $\mc{A}^4_{clr}$ is
\begin{align*}
\{\varepsilon\}d_{01}\{1\}d_{13}     \left\{\begin{matrix} 1 \\ 1\\1\end{matrix}\right\}         \lma \delta(a)\\ \delta(a)\\ \delta(a) \rma           \left\{\begin{matrix} 2 \\ 3\\4\end{matrix}\right\}         &\lma e\\ d_{12}\\ e \rma      \left\{\begin{matrix} 2 \\ 3\\3\\4\end{matrix}\right\}                  \lma e\\\delta(a)\\ \delta(a)\\e\rma            \left\{\begin{matrix} 2 \\ 2\\4\\4\end{matrix}\right\}       \\ & \lma d_{21}\\ d_{21}\rma          \left\{\begin{matrix} 2 \\ 4\end{matrix}\right\}      \lma  \delta(a)\\ e \rma         \left\{\begin{matrix} 4 \\ 4\end{matrix}\right\} d_{21}\{4\} \, d_{10}\{\varepsilon\}
\end{align*}
and the corresponding $4$-coloring that is obtained by the state distribution of this transition is
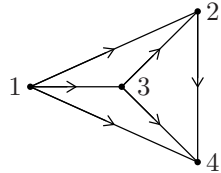
\bc
\begin{texdraw}
\drawdim mm \linewd 0.21 \arrowheadtype t:H \arrowheadsize l:1.2
w:1.4

\move(42 10)\fcir f:0 r:0.4 \lvec
(32 20) \move (32 20) \avec (37 15)  \move(42 30)\fcir f:0 r:0.4 \lvec
(32 20)  \move (32 20) \avec (37 25) \move(42 30)  \avec (42 20)\move (42 30)\lvec
(42 10) \move(20 20)\fcir f:0 r:0.4  \avec (26 20)\move (20 20)\lvec
(32 20) \fcir f:0 r:0.4  \move (20 20) \lvec (42 30) \move (20 20) \avec (31 25)
\move (20 20) \lvec (42 10) \move (20 20) \avec (31 15)

\textref h:C v:C   \textref h:C v:C   \htext(18 20){$1$} \htext(35 20){$3$} \htext(44 10){$4$} \htext(44 30){$2$}
\end{texdraw}
\ec
Note that the elementary graph $\Pi$ is not necessary for the representation of $G$ and $F$. More generally, all planar graphs can be represented without employing $\Pi$. On the other hand, the non-planar graph $K_{3,3}$
is expressed as
\[
K_{3,3}=\lma I_{01}\\ I_{01}\\I_{01}\rma \lma I_{13}\\ I_{13}\\I_{13}\rma \lma a\\ \vdots \\ a\rma \Pi _s \lma I_{31}\\ I_{31}\\I_{31}\rma \lma I_{10}\\ I_{10}\\I_{10}\rma
\]
where in the third parenthesis there are $9$ $a$'s and $\Pi_s$ stands for the graph that is associated with the permutation
\[
\lma 1 & 2 & 3 & 4 & 5 & 6 & 7& 8 & 9 \\ 1 & 4 & 7 & 2 & 5 & 8 & 3& 6 & 9 \rma
\]
Notice that as it is shown in \cite{BK1} for every permutation we can construct, inductively by $\Pi$ and $E$ a graph that represents it. The graph $K_{3,3}$ is $2$-colorable and an accepting transition of $\mc{A}^2_{clr}$ is
\begin{align*}
\{\varepsilon\}\lma d_{01}\\ d_{01}\\d_{01}\rma   \left\{\begin{matrix} 1 \\ 1\\1\end{matrix}\right\}  \lma d_{13}\\ d_{13}\\d_{13}\rma  \left\{\begin{matrix} 1 \\ \vdots \\1\end{matrix}\right\}  \lma \delta(a)\\ \vdots \\ \delta(a)\rma \left\{\begin{matrix} 2 \\ \vdots \\2\end{matrix}\right\}  d _s \left\{\begin{matrix} 2 \\ \vdots \\2\end{matrix}\right\}  \lma d_{31}\\ d_{31}\\d_{31}\rma  \left\{\begin{matrix} 2 \\ 2 \\2\end{matrix}\right\}  \lma d_{10}\\ d_{10}\\d_{10}\rma  \{\varepsilon\}
\end{align*}
Hence, if we denote by $kCol$ the set of all $k$-colorable graphs, we obtain
\begin{theorem}
For every $k\in \N$, the graph language $kCol$ is   recognizable.
\end{theorem}
\section{Conclusion and Future Work}

It is shown that $k$-colorability is graph automaton recognizable even when the automaton operates over the most trivial abelian relational graphoid. This indicates that the graph automaton is a robust recognition mechanism and evokes numerous issues regarding the class of automaton recognizable graph languages.
\begin{itemize}
  \item The introduced graph automata are nondeterministic in the sense that if $\sigma\in \Sigma _{m,n}$ then $\delta (\sigma)$ is a relation in $Q^m\times Q^n$. The deterministic version of this definition is obtained by requiring that $\delta (\sigma)$ is a function from $Q^m$ to $Q^n$, i.e., $\delta (\sigma)$ is an element of $Funct (Q)$ which is a submagmoid of $Rel(Q)$ called the magmoid of functions \cite{BK3}. It is interesting to compare the two classes and in particular to investigate the existence of a deterministic graph automaton recognizing $k$-colorable graphs.
  \item In \cite{BK3} it is shown that the time complexity for checking if a specific graph belongs to the behavior of a graph automaton grows polynomially as a function of the number of states $|Q|$. It is important to calculate the complexity of the membership problem for a given graph automaton as a function of the size of the graph for both the deterministic and the nondeterministic case. Such a result would classify automaton recognizable graph languages according to their computational complexity.
  \item In \cite{Kal2} it is proved that an infinite number of non-isomorphic abelian relational graphoids exists. Two questions that naturally arise  concern  the recognition capacity of the corresponding automata as well as the existence of non-abelian or non-relational graphoids.
\end{itemize}

\end{document}